\documentclass[12pt,epsf]{article}
\usepackage{graphicx}
\usepackage{epsf}
\begin{document}

\def\a{\alpha}
\def\b{\beta}
\def\c{\varepsilon}
\def\d{\delta}
\def\e{\epsilon}
\def\f{\phi}
\def\g{\gamma}
\def\h{\theta}
\def\k{\kappa}
\def\l{\lambda}
\def\m{\mu}
\def\n{\nu}
\def\p{\psi}
\def\q{\partial}
\def\r{\rho}
\def\s{\sigma}
\def\t{\tau}
\def\u{\upsilon}
\def\v{\varphi}
\def\w{\omega}
\def\x{\xi}
\def\y{\eta}
\def\z{\zeta}
\def\D{\Delta}
\def\G{\Gamma}
\def\H{\Theta}
\def\L{\Lambda}
\def\F{\Phi}
\def\P{\Psi}
\def\S{\Sigma}

\def\o{\over}
\def\beq{\begin{eqnarray}}
\def\eeq{\end{eqnarray}}
\newcommand{\gsim}{ \mathop{}_{\textstyle \sim}^{\textstyle >} }
\newcommand{\lsim}{ \mathop{}_{\textstyle \sim}^{\textstyle <} }
\newcommand{\vev}[1]{ \left\langle {#1} \right\rangle }
\newcommand{\bra}[1]{ \langle {#1} | }
\newcommand{\ket}[1]{ | {#1} \rangle }
\newcommand{\EV}{ {\rm eV} }
\newcommand{\KEV}{ {\rm keV} }
\newcommand{\MEV}{ {\rm MeV} }
\newcommand{\GEV}{ {\rm GeV} }
\newcommand{\TEV}{ {\rm TeV} }
\def\diag{\mathop{\rm diag}\nolimits}
\def\Spin{\mathop{\rm Spin}}
\def\SO{\mathop{\rm SO}}
\def\O{\mathop{\rm O}}
\def\SU{\mathop{\rm SU}}
\def\U{\mathop{\rm U}}
\def\Sp{\mathop{\rm Sp}}
\def\SL{\mathop{\rm SL}}
\def\tr{\mathop{\rm tr}}

\def\IJMP{Int.~J.~Mod.~Phys. }
\def\MPL{Mod.~Phys.~Lett. }
\def\NP{Nucl.~Phys. }
\def\PL{Phys.~Lett. }
\def\PR{Phys.~Rev. }
\def\PRL{Phys.~Rev.~Lett. }
\def\PTP{Prog.~Theor.~Phys. }
\def\ZP{Z.~Phys. }

\newcommand{\gtrsim}{ \mathop{}_{\textstyle \sim}^{\textstyle >} }
\newcommand{\lesssim}{ \mathop{}_{\textstyle \sim}^{\textstyle <} }
\newcommand{\rem}[1]{{\bf #1}}


\baselineskip 0.7cm

\begin{titlepage}

\def\thefootnote{\fnsymbol{footnote}}

\begin{center}

\hfill TU-810\\
\hfill IPMU 08-0009\\
\hfill February, 2008\\

\vskip .5in

{\Large \bf 
Test of Anomaly Mediation at the LHC
}

\vskip .5in

{\large
$^{(a)}$Shoji Asai, $^{(b)}$Takeo Moroi and $^{(a,c)}$T.T. Yanagida
}

\vskip 0.25in

{\em $^{(a)}$Department of Physics, University of Tokyo,
Tokyo 113-0033, Japan
}

\vskip 0.1in

{\em $^{(b)}$Department of Physics, Tohoku University,
Sendai 980-8578, Japan}

\vskip 0.1in

{\em $^{(c)}$Institute for the Physics and Mathematics of the Universe,\\
University of Tokyo, Chiba 277-8568, Japan}

\end{center}
\vskip .5in

\begin{abstract}

  In the anomaly-mediated supersymmetry breaking model with the
  assumption of the generic form of K\"ahler potential, gauginos are
  the only kinematically accessible superparticles to the LHC.  We
  consider the LHC phenomenology of such a model assuming that the
  gluino is lighter than 1 TeV.  We show that a significant number of
  charged Winos, which may travel $O(10\ {\rm cm})$ before the decay,
  are produced from the gluino production processes.  Thus, in this
  class of model, it will be very important to search for short
  charged tracks using inner detectors.  We also show that, if a large
  number of the charged Wino tracks are identified, the lifetime of
  the charged Wino can be measured, which provides us a test of
  anomaly mediation.

\end{abstract}

\end{titlepage}

\section{Introduction}

Once supersymmetry (SUSY) is broken in a hidden sector, the
SUSY-breaking effects are automatically transferred to the SUSY
standard model (SSM) sector by the tree-level supergarvity (SUGRA)
interactions.  That is, squarks and sleptons acquire soft
SUSY-breaking masses of order of the gravitino masses $m_{3/2}$.
However, the gauginos are all massless at this level and hence one
usually introduces couplings between a hidden field $Z$ and the
gauge-field-strength superfield $W^i_\alpha$ as $(Z/M_{\rm
  PL})W^{i\alpha} W^i_\alpha$. Here, $M_{\rm PL}\simeq 2.4\times
10^{18}$ GeV is the reduced Planck mass and the hidden field $Z$ is
supposed to have a SUSY-breaking $F$ term. Provided that the $F$ term
is the dominant component of the SUSY breaking, one obtains the
gaugino masses of order of the gravitino mass $m_{3/2}$. In this
scenario the hidden field $Z$ must be completely neutral to have the
above coupling to the gauge kinetic functions. However, the neutrality
of the hidden field $Z$ causes serious cosmological problems; the
so-called modulus problem \cite{Ibe:2006ck} and the over-production of
the gravitinos in inflaton decays \cite{GravFromInflaton}.

It was sometimes ago pointed out \cite{Giudice:1998xp,Randall:1998uk}
that the one-loop quantum effects induce the gaugino masses in the SSM
without the neutral hidden field. This mechanism is called as "anomaly
mediation".  Namely, the anomaly mediation always takes place in the
quantum SUGRA and the gauginos become massive without any neutral
hidden field once the SUSY is broken. Therefore, the model is free
from the above cosmological problems since there is no need to
introduce a neutral hidden field.

The anomaly mediation predicts so-called split SUSY spectrum where
squarks and sleptons may have masses of the order 100 TeV while the
masses of gauginos are in the range of 100 GeV $-$ 1 TeV. (The gaugino
masses is suppressed, since they are generated at the one-loop level.)
Because of the relatively large masses of squarks we need a very
precious fine-tuning of parameters to obtain the correct electro-weak
symmetry breaking, but on the other hand it solves many serious
problems in the SSM. First of all the flavor-changing neutral current
and CP-violation problems become very milder due to the large masses
of squarks and sleptons.  We may naturally explain no discovery of
Higgs at LEP and no discovery of proton decays induced by
dimension-five operators. Furthermore, the gravitino mass is also
predicted at the order of 100 TeV, which makes another cosmological
gravitino problems much less severe \cite{GravitinoBBN}. (The
gravitinos are also produced by particle scatterings in the thermal
bath in early universe. We have upper bounds of the reheating
temperature $T_R$ to avoid over-production of the gravitinos depending
on the gravitino mass.)  In fact, it has been pointed out
\cite{Ibe:2004tg} that the leptogenesis does work in the
anomaly-mediation model, since the reheating temperature $T_R$ can be
as high as $10^{10}$ GeV without any confliction with cosmology.

Although the squarks and sleptons are so heavy as explained above, the
masses of gauginos may be in the accessible range to the LHC
experiments. The anomaly mediation predicts a certain relation among
the gaugino masses such as $M_2<M_1<M_3$ (where $M_1$, $M_2$, and
$M_3$ are gaugino masses for $U(1)_{\rm Y}$, $SU(2)_{\rm L}$, and
$SU(3)_{\rm C}$ gauge groups, respectively) in a large region of the
parameter space \cite{Ibe:2006de}. And the charged wino has a
considerably long lifetime as $c\tau_{\tilde{W}^\pm}\sim 5\ {\rm cm}$
(with $c$ being the speed of light).  Interesting enough is that the
lifetime is almost independent of the parameters in the
anomaly-mediation model. Thus, we consider that not only the
observation of such a long-lived charged particle but also the
measurement of its lifetime at the LHC provide a serious test of the
anomaly-mediation model.

In the previous work \cite{Ibe:2006de} we consider a very pessimistic
situation where the gluino $\tilde{g}$ is too heavy to be produced at
the LHC and hence the number of produced winos $\tilde{W}$ is limited.
In this letter we assume that the gluino is lighter than 1 TeV and
they are efficiently produced at the LHC. And we show that the large
number of winos are produced through the gluino production and the
lifetime of the charged wino may be measured at the LHC to test the
anomaly mediation.

We also note that it will be difficult to confirm supersymmetry even
at the LHC experiment, if the present anomaly-mediation model is
realized.  Thus, it is important to see if the properties of the
unstable charged particle, observed as short charged tracks, are
consistent with the prediction of the anomaly-mediation model.

\section{Properties of Gauginos}

First, we summarize the mass spectrum of superparticles in our
analysis.  In the anomaly-mediation model where the Higgsinos (as well
as sfermions) acquire masses of $O(10\ {\rm TeV})$, radiative
correction due to the Higgs-Higgsino loop diagram significantly
changes the gaugino-mass relation predicted in the pure
anomaly-mediation model \cite{Gherghetta:1999sw}.  In our study, we
consider the case $M_{2}<M_{1}<M_{3}$, but the ratio $M_{3}/M_{2}$ is
assumed to be much smaller than the predicted value from the pure
anomaly-mediation model.  In our following numerical study, we use the
gaugino mass parameters
\begin{eqnarray}
  M_2 = 200\ {\rm GeV}, ~~~M_3 = 1\ {\rm TeV},
  \label{M_23}
\end{eqnarray}
and consider the case where gluino dominantly decays as
$\tilde{g}\rightarrow\tilde{W}q\bar{q}$ (and hence the bino is
irrelevant in our study).  In this case, a large number of winos are
produced from the gluino production processes at the LHC, contrary to
the assumption used in \cite{Ibe:2006de}.  Then, detailed studies of
the properties of wino may be possible.  For example, with the gluino
mass of $1\ {\rm TeV}$, the gluino production cross section at the LHC
is $\sigma_{pp\rightarrow\tilde{g}\tilde{g}}\simeq 220\ {\rm fb}$.
(Here, we have taken the renormalization scale to be the gluino mass.)
Since the gluino decays into the charged wino with the branching ratio of
$2/3$, charged winos of $O(10^5)$ is produced with the Luminosity of
$100\ {\rm fb}^{-1}$ (although many of them decay without being
detected).

Next, let us discuss the properties of winos in the anomaly-mediation
model.  The mass difference between charged and neutral winos
originates dominantly from 1-loop Feynmann diagrams with electro-weak
bosons in the loop.  Then, charged wino becomes heavier than neutral
one, and the mass difference is given by \cite{Cheng:1998hc}
\begin{eqnarray}
  \delta m_{\tilde{W}} =  m_{{\tilde W}^{\pm}}-m_{{\tilde W}^{0}}
  = \frac{g_{2}^{2}}{16\pi^{2}} M_{2} 
  \left[ f(r_{W})-\cos^{2}\h_{W} f(r_{Z})-\sin^{2}\h_{W} f(0) \right],
  \label{deltaM}
\end{eqnarray}
where $f(r)= \int^{1}_{0} dx(2 + 2 x^{2}) \ln[x^{2} + (1-x)r^{2}]$ and
$r_{i}=m_{i}/M_{2}$.  The mass difference is in the range $155\ {\rm
  MeV}\lesssim \delta m_{\tilde{W}}\lesssim 170\ {\rm MeV}$, which is
much smaller than the (expected) wino mass parameter $M_2$.  If
$\tilde{W}^0$ is the LSP, $\tilde{W}^\pm$ dominantly decays as
$\tilde{W}^\pm\rightarrow\tilde{W}^0\pi^\pm$ and its lifetime becomes
very long; in such a case, the lifetime of $\tilde{W}^\pm$ is given by
\begin{eqnarray}
  \tau_{\tilde{W}^\pm}^{-1} = 
  \frac{2 G_{F}^{2}}{\pi} \cos^{2}\h_{c} f_{\pi}^{2}
  \delta m_{\tilde{W}}^{3} 
  \left(1- \frac{m_{\pi}^{2}}{\delta m_{\tilde{W}}^{2}}\right)^{1/2},
  \label{WinoWidth}
\end{eqnarray}
where $f_{\pi} \simeq 130$\,MeV, and $\h_{c}$ is the Cabbibo angle.
Numerically, the lifetime is of ${\cal O}(10^{-10}\ {\rm sec})$.
Importantly, the mass difference $\delta m_{\tilde{W}}$, and hence the
lifetime $\tau_{\tilde{W}^\pm}$, are insensitive to the wino mass
parameter $M_2$.  Thus, if the lifetime of charged wino is
experimentally determined, it provides an important test
of the anomaly-mediation model.

The smallness of the mass difference has a significant implication for
collider experiments when the neutral wino $\tilde{W}^0$ is the LSP.
In this case, the NLSP, charged wino $\tilde{W}^\pm$, decays into the
neutral one $\tilde{W}^0$ by emitting very soft $\pi^\pm$ which easily
escape the detection.  This fact makes the discovery of
$\tilde{W}^\pm$ at the LHC very challenging.  However, with the
lifetime estimated above, the charged winos produced in the LHC
experiment are expected to travel $\sim {\cal O}(1-10\ {\rm cm})$
before they decay.  Thus, they travel for macroscopic distances and
some of them may decay after traveling through some of the detectors.
In such a case, charged winos may be observed as energetic short
charged tracks.

\begin{figure}[t]
  \centerline{\epsfxsize=0.5\textwidth\epsfbox{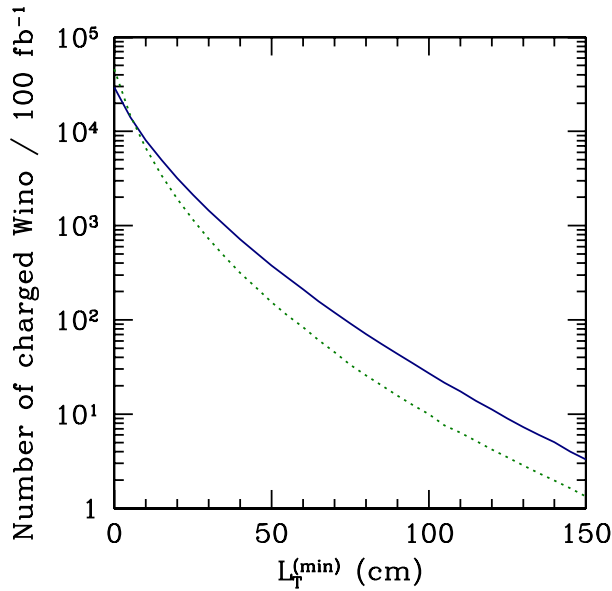}}
  \caption{Number of charged winos produced at the LHC, which travel
    transverse distance longer than $L_{\rm T}^{\rm (min)}$.  The
    solid line is for the process $pp\rightarrow\tilde{g}\tilde{g}$,
    while the dotted line is for
    $pp\rightarrow\tilde{W}\tilde{W}j_{p_{\rm T}>100\ {\rm GeV}}$
    Here, we use the luminosity of $100\ {\rm fb}^{-1}$.}
 \label{fig:cs}
\end{figure}

In Fig.\ \ref{fig:cs}, we plot the number of $\tilde{W}^\pm$ from the
gluino production process $pp\rightarrow\tilde{g}\tilde{g}$, requiring
that the transverse travel length of $\tilde{W}^\pm$ be longer than
than $L_{\rm T}^{\rm (min)}$ (with the luminosity of $100\ {\rm
  fb}^{-1}$).  There are still a few hundred of the non-decay winos at
even $L_{\rm T}^{\rm (min)}=54\ {\rm cm}$, at which the transition
radiation tracker (TRT) is set in the ATLAS detector.  By using these
samples, we may be able to extract information on the winos as we will
discuss in the following section.  This may provide quantitative tests
of the anomaly-mediation model.  For comparison, we also plot the
number of $\tilde{W}^\pm$ from the Drell-Yan induced process
$pp\rightarrow\tilde{W}^\pm\tilde{W}^\mp j_{p_{\rm T}>100\ {\rm GeV}}$
and $pp\rightarrow\tilde{W}^0\tilde{W}^\mp j_{p_{\rm T}>100\ {\rm
    GeV}}$, where $j_{p_{\rm T}>100\ {\rm GeV}}$ denotes the energetic
jet with the transverse momentum larger than $100\ {\rm GeV}$.  (The
jet is a necessary trigger for the event \cite{Feng:1999fu}.)  With
the mass spectrum we have adopted, the gluino production process
produces a larger amount of charged winos than the wino + jet
production process if $L_{\rm T}^{\rm (min)}\gtrsim 10\ {\rm cm}$.  In
the following numerical analysis, for simplicity, we only consider the
winos produced by the process $pp\rightarrow\tilde{g}\tilde{g}$.

\section{Measurement of the Lifetime}

Now, we are at the position to discuss LHC phenomenology with the wino
LSP.  Since we are interested in the charged wino whose lifetime is
$O(10^{-10}\ {\rm sec})$, most of the charged winos produced at the
LHC experiment will decay inside the detectors.  Thus, we should look
for short charged tracks with high momentum which disappear inside the
detector.  If we can obtain a large number of samples of short charged
tracks, we may check if the properties of the observed short-lived
charged particle are consistent with the prediction of
anomaly-mediation model.

However, in the actual situation, such a study will be non-trivial.
This is because, first of all, it will be challenging to find such
short charged tracks and, second, accurate measurements of the travel
length should be also non-trivial.

Importantly, the ATLAS detector has the TRT which may be useful for
the detailed study of the properties of charged wino.  The TRT is
located at 54 -- 107 cm from the beam axis \cite{ATLASin} and
continuously follows charged tracks.  The TRT may be used to find
charged-wino tracks.  In the following, we show how well we can study
properties of charged wino, assuming that the charged-wino tracks can
be found with high efficiency with the TRT.  We have checked the
tracking efficiencies and precision of the track resolution for the
various decay position in TRT, and they are found to be stable.  So
this assumption is reasonable.

First, as discussed in \cite{Asai:2007sw}, once the wino tracks are
found, wino mass can be determined by combining the time-of-flight
information with the momentum information. The resolution of the
velocity $\beta$ is about 0.1 if $\beta<0.85$.  Then, the mass can be
determined with the accuracy of 10\% if enough samples of the exotic
tracks are available.

If a large number of samples of the charged wino are obtained in the
form of short charged tracks, distribution of the length of those
tracks $L$ can be derived.  From the distribution of $L$, the lifetime
of charged wino may be also determined.  As we have mentioned, the
lifetime of charged wino in the anomaly-mediated model is accurately
calculated, an interesting test of the anomaly-mediation model is
possible by comparing the experimentally determined lifetime with the
theoretical prediction.

If a large number of charged winos {\it at rest} are available, the
number of $\tilde{W}^\pm$ decreases with time as
$N_{\tilde{W}^\pm}\propto e^{-t/\tau_{\tilde{W}^\pm}}$.  Thus, by
fitting the survival probability with the exponential function, one
may be able to determine the lifetime.  However, in the LHC
experiment, charged winos are produced with various velocities, and
hence we have to take into account the effect of $\gamma$-factor.  In
addition, probably the precise determination of the travel length is
possible only if (i) charged wino travels some amount of length in the
TRT, and (ii) charged wino decays inside the TRT.

In the following discussion, let us consider how we can determine the
lifetime by using only such limited samples.  For this purpose, we
assume that the charged winos satisfying (i) and (ii) can be
identified by the off-line analysis and that their travel length can
be determined.

In order to quantize the conditions (i) and (ii), we define the
transverse travel length $L_{\rm T}$ as
\begin{eqnarray}
  L_{\rm T} = L \sin\theta,
\end{eqnarray}
where $\theta$ is the direction of the charged wino with respect to
the beam axis.  Then, the conditions (i) and (ii) becomes that the
$L_{\rm T}$ is in the ``fiducial'' volume of the TRT; $L_{\rm T}^{\rm
  (min)}<L_{\rm T}<L_{\rm T}^{\rm (max)}$ (with some relevant
constraint on $\theta$).  In our numerical study, we take $L_{\rm
  T}^{\rm (min)}=60\ {\rm cm}$ and $L_{\rm T}^{\rm (max)}=100\ {\rm
  cm}$.

Using the fact that the momentum of individual charged track is
measurable, we parametrize the observable constructed from $L_{\rm
T}$ and the momentum as
\begin{eqnarray}
  \frac{L_{\rm T} - L_{\rm T}^{\rm (min)}}{|{\bf p}_{\rm T}|} \equiv
  t_{\rm D} m_{\tilde{W}^\pm}^{-1},
\end{eqnarray}
where ${\bf p}_{\rm T}$ is the transverse momentum, i.e., $|{\bf
p}_{\rm T}|=|{\bf p}|\sin\theta$.  The physical meaning of $t_{\rm D}$
is the time interval in the rest frame of $\tilde{W}^\pm$ between the
moment corresponding to $L_{\rm T}=L_{\rm T}^{\rm (min)}$ and that of
the decay.  Importantly, when $L_T^{\rm (max)}\rightarrow\infty$, the
distribution of the variable $t_{\rm D}$ obeys $P(t_{\rm
D})=P_0e^{-t_{\rm D}/\tau_{\tilde{W}^\pm}}$, where $P_0$ is the
normalization constant.  Thus, from the distribution of the variable
$t_{\rm D}m_{\tilde{W}^\pm}^{-1}$, we obtain information about the
combination $\tau_{\tilde{W}^\pm}m_{\tilde{W}^\pm}^{-1}$.

\begin{figure}[t]
  \centerline{\epsfxsize=0.5\textwidth\epsfbox{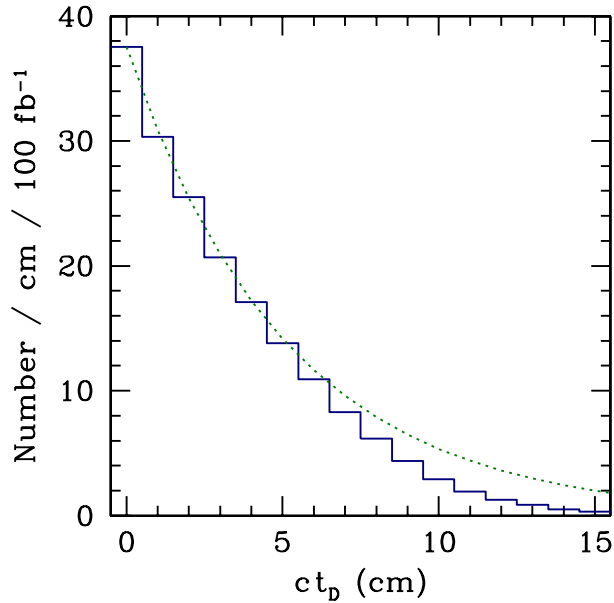}}
  \caption{The histogram is the distribution of the variable $t_{\rm
      D}$ multiplied by the speed of light.  The dotted line is the
    number of events expected from the exponential distribution (i.e.,
    in the case of $L_{\rm T}^{\rm (max)}\rightarrow\infty$).}
 \label{fig:ctdist}
\end{figure}

In the actual situation, $L_T^{\rm (max)}$ is finite, and hence the
distribution $P(t_{\rm D})$ does not exactly follow the exponential
behavior.  However, even in that case, we may be able to constrain
$\tau_{\tilde{W}^\pm}m_{\tilde{W}^\pm}^{-1}$.  In Fig.\
\ref{fig:ctdist}, we plot the distribution of $t_{\rm D}$ for $L_{\rm
  T}^{\rm (max)}=100\ {\rm cm}$.  Here, using MadGraph/MadEvent
package \cite{MGME}, we have generated $10^7$ signal events and
calculated the distribution with Monte Carlo analysis.  In addition,
in the same figure, we show the distribution for the case $L_{\rm
  T}^{\rm (max)}\rightarrow\infty$.  As one can see, for $ct_{\rm
  D}\lesssim 10\ {\rm cm}$, $P(t_{\rm D})$ is well approximated by the
exponential function.  Thus, if the distribution of $t_{\rm D}$ is
obtained, we may be able to extract the information about the lifetime
of the charged wino.

\begin{figure}[t]
  \centerline{\epsfxsize=0.5\textwidth\epsfbox{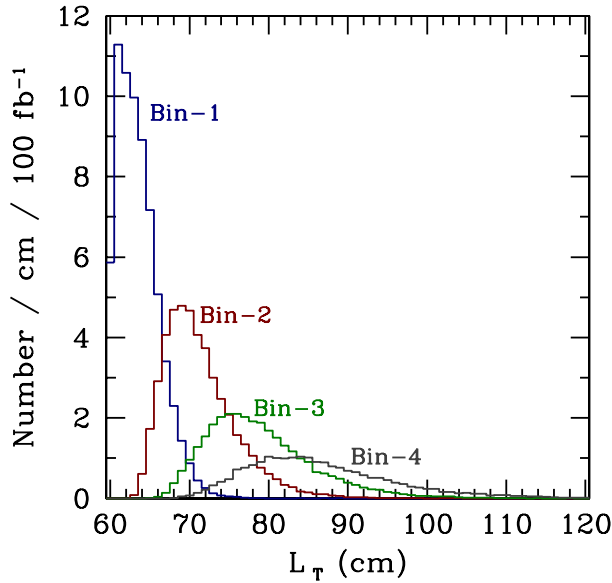}}
  \caption{Distributions of the variable $L_{\rm T}$ for samples in
    bins $1-4$.  Here, the luminosity of $100\ {\rm fb}^{-1}$ is 
    used.}
  \label{fig:ltdist}
\end{figure}

In order to see how well the lifetime (more accurately, the
combination $\tau_{\tilde{W}^\pm}m_{\tilde{W}^\pm}^{-1}$) can be
constrained, we first generate event samples using the underlying
parameters given in (\ref{M_23}) for the luminosity of $100\ {\rm
  fb}^{-1}$ and calculate the distribution $t_{\rm D}$, $P(t_{\rm
  D})$.  For the statistical analysis, we classify the events into
four bins, which are
\begin{itemize}
\item[1.] $0\ {\rm cm} < ct_{\rm D}\leq 2\ {\rm cm}$
($0\ {\rm cm/GeV}< ct_{\rm D}m_{\tilde{W}^\pm}^{-1}\leq 0.01\ 
{\rm cm/GeV}$),
\item[2.] $2\ {\rm cm}<ct_{\rm D}\leq 4\ {\rm cm}$
($0.01\ {\rm cm/GeV}<ct_{\rm D}m_{\tilde{W}^\pm}^{-1}\leq 0.02\ 
{\rm cm/GeV}$),
\item[3.] $4\ {\rm cm}<ct_{\rm D}\leq 6\ {\rm cm}$
($0.02\ {\rm cm/GeV}<ct_{\rm D}m_{\tilde{W}^\pm}^{-1}\leq 0.03\ 
{\rm cm/GeV}$),
\item[4.] $6\ {\rm cm}<ct_{\rm D}\leq 8\ {\rm cm}$
($0.03\ {\rm cm/GeV}<ct_{\rm D}m_{\tilde{W}^\pm}^{-1}\leq 0.04\ 
{\rm cm/GeV}$).
\end{itemize}
We denote the number of events in each bin as $n_k$ ($k=1-4$).
Distributions of the variable $L_{\rm T}$ for samples in bins $1-4$
are shown in Fig.\ \ref{fig:ltdist}.  We can see that most of the
charged winos in the sample events decay within the TRT.
Then, we compare $n_k$ ($k=1-4$) with the expected number of events
estimated with the postulated lifetime $\tau$.  Expected number of
events in four bins are denoted as $\bar{n}_k (\tau)\propto
e^{-kd_{\rm bin}/c\tau} -e^{-(k+1)d_{\rm bin}/c\tau}$, where $d_{\rm
bin}$ is the width of the bins and is $d_{\rm bin}=2\ {\rm cm}$.
With $n_k$ and $\bar{n}_k$, we calculate the $\chi^2$ variable.  In
order to use only the shape information of the distribution, we treat
the normalization as a free parameter, and the $\chi^2$ variable is
defined as
\begin{eqnarray}
  \chi^2 (\tau) 
  = \sum_k \frac{\left[ n_k - N_0 \bar{n}_k (\tau) \right]^2}{n_k}
  = \sum_k n_k - 
  \frac{\left[ \sum_k \bar{n}_k (\tau) \right]^2}
  {\sum_k  \left[ \bar{n}_k^2(\tau) /n_k \right] },
\end{eqnarray}
where $N_0$ is the normalization constant which minimizes $\chi^2$.
The $\chi^2$ variable obtained above fluctuates, depending on the set
of event samples $\{n_k\}$.  In order to estimate the typical
uncertainty in the determination of the lifetime, we repeat the above
process to obtain the averaged value of $\chi^2(\tau)$ for each
postulated value of $\tau$.

\begin{figure}[t]
  \centerline{\epsfxsize=0.5\textwidth\epsfbox{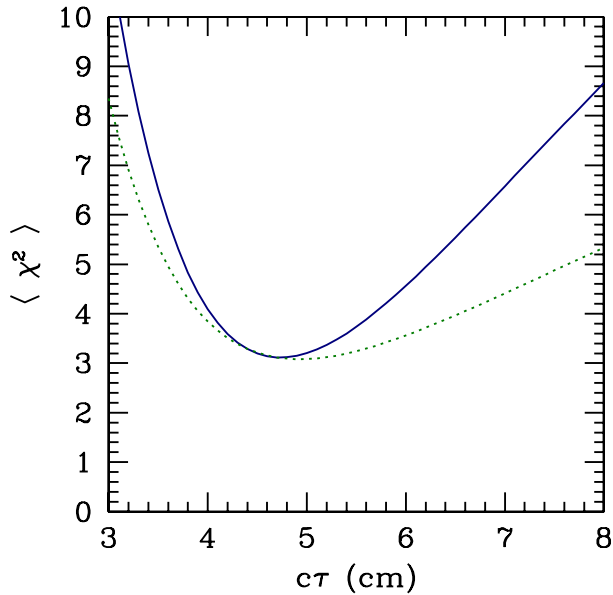}}
  \caption{The averaged value of the $\chi^2$ variable as a function
    of the postulated value of the lifetime $\tau$.  Here, the
    $\chi^2$ variable is calculated with four bins with $d_{\rm
    bin}=2\ {\rm cm}$ (solid) and $1.5\ {\rm cm}$ (dotted) with the
    luminosity of $100\ {\rm fb}^{-1}$.  The input value of the
    lifetime of the charged wino is $c\tau_{\tilde{W}^\pm} =5.1\ {\rm
    cm}$.}
    \label{fig:chi2bar}
\end{figure}

The averaged value of $\chi^2(\tau)$ is shown in Fig.\
\ref{fig:chi2bar}.  The typical constraint on the lifetime is
estimated from
$\delta\langle\chi^2(\tau)\rangle\equiv\langle\chi^2(\tau)\rangle
-\langle\chi^2\rangle_{\rm min}<1$ where $\langle\chi^2\rangle_{\rm
min}$ is the minimum value of $\langle\chi^2(\tau)\rangle$.  (Notice
that $\langle\chi^2\rangle_{\rm min}\simeq 3$, which is consistent
with the expected value of the $\chi^2$ variable with three degrees of
freedom.)  Then, we obtain the constraint $4.0\ {\rm cm}<c\tau<5.7{\rm
cm}$ (or $0.020\ {\rm cm/GeV}<c\tau m_{\tilde{W}^\pm}^{-1}<0.029{\rm
cm/GeV}$, if no information about the wino mass is available).  Thus,
the lifetime is constrained with the uncertainty of about $10-20\ \%$.

The best-fit value of the lifetime, which is $c\tau_{\rm best}\simeq
4.8{\rm cm}$, is smaller than the input value of the lifetime.  This
is probably due to the fact that the number of events in higher bins
(in particular, that of 4th bin) are smaller than the expectation
values from the exponential distribution because only the events with
$L_{\rm T}<100\ {\rm cm}$ are adopted (see Fig.\ \ref{fig:ctdist}).
If we limit ourselves to the samples with smaller value of $t_{\rm
  D}$, the best-fit value of the lifetime becomes closer to the input
value.  To see this, we also calculate the averaged value of the
$\chi^2$ variable with the four bins with the interval of $1.5\ {\rm
  cm}$, instead of $2\ {\rm cm}$, and the result is also shown in
Fig.\ \ref{fig:chi2bar}.  In this case, the best-fit value becomes
$5.0\ {\rm cm}$ and is closer to the input value.  However, since the
number of events decreases in this case, the uncertainty becomes
larger.

\begin{figure}[t]
  \centerline{\epsfxsize=0.5\textwidth\epsfbox{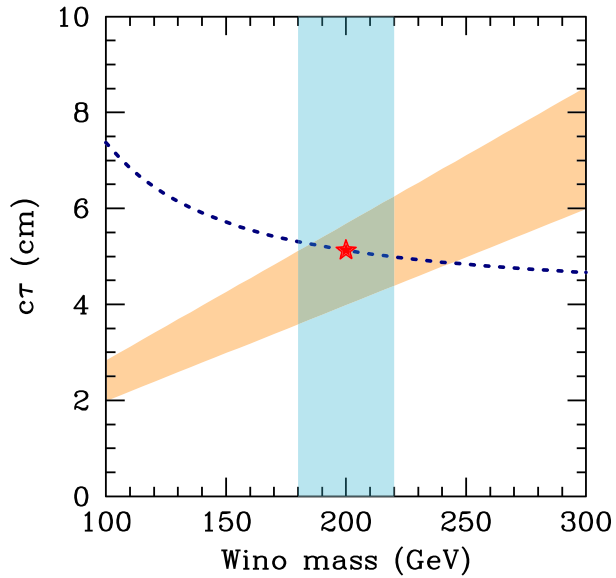}}
  \caption{Expected experimental and theoretical constraints on the 
    wino mass vs.\ $c\tau$ plane.  The vertical band is the constraint on 
    the wino mass using the velocity and momentum information, while  
    another band is from the bound on the wino lifetime as a function of
    the wino mass.  The dotted line is the theoretical prediction of 
    the lifetime of the wino (multiplied by the speed of light).  The star
    at the center is the input point.}
    \label{fig:constall}
\end{figure}

Finally, in Fig.\ \ref{fig:constall}, we summarize expected
experimental constraints on the wino mass vs.\ $c\tau$ plane.  In the
same figure, we also plot the theoretical prediction on the lifetime
of the wino as a function of the wino mass.  We can see that, if the
(short) charged wino tracks can be identified with a sizable
efficiency, we can provide an interesting test of the
anomaly-mediation model.

\section{Conclusions}

In this letter, we have proposed a new procedure to test the
anomaly-mediation model in which only gauginos are kinematically
accessible to the LHC.  We have paid particular attention to the study
of the charged Wino, which is expected to behave as a short charged
track in the detectors.  As we have discussed, the mass of the charged
wino may be determined from the velocity and momentum information
while the information about the lifetime of the charged wino may be
obtained from the distribution of the travel length.  Combining these
two, non-trivial test of the anomaly-mediation model may be possible.
Thus, we emphasize the importance of the search and the study of short
charged tracks using the TRT (and other detector components).

Here, we have concentrated on informations available from the charged
Wino tracks observed by the TRT.  However, constraints on the gaugino
masses (in particular, those on the wino mass) is also obtained from
the invariant-mass distribution of the jets produced by the decay
process of the gluino: $\tilde{g}\rightarrow\tilde{W}q\bar{q}$
\cite{Asai:2007sw}.  Such informations also provide important and
independent test of the anomaly-mediation model.

In this letter, possibility of determining the lifetime of the charged
wino has been discussed.  However, we have not included any detailed
detector effects in the present analysis.  In addition, we have assumed
that the short charged track can be easily identified irrespective of
$L_{\rm T}$.  In the realistic situation, however, more detailed
understanding of the efficiency to discover the short charged tracks is
necessary.  Such a study will be performed elsewhere \cite{AsaMorYan}.

{\it Acknowledgement}: This work was supported in part by the
Grant-in-Aid for Scientific Research from the Ministry of Education,
Science, Sports, and Culture of Japan, No.\ 19540255 (T.M.).

\end{document}